\documentclass[aps,prd,twocolumn,groupedaddress,nofootinbib]{revtex4-2}

\usepackage{amsmath,amssymb,amsthm,graphicx}
\usepackage[dvipsnames]{xcolor}
\usepackage{multirow}
\usepackage{adjustbox}
\usepackage{comment}
\usepackage{enumerate}
\usepackage{hyperref}
\usepackage{cleveref}
\usepackage{bm}
\usepackage{subcaption}
\usepackage{xcolor}

\newcommand{\E}{\mathcal{E}}
\newcommand{\gij}{g_{ij'}}

\begin{document}

\title{Multi-scale Optimal Transport for Complete Collider Events}

\author{Tianji Cai}
\email{tianjiresearch@gmail.com}
\affiliation{School of Physical Science and Engineering, Tongji University, Shanghai 200092, China}
\affiliation{State Key Laboratory of Autonomous Intelligent Unmanned Systems, MOE Frontiers Science Center for Intelligent Autonomous Systems, Tongji University, Shanghai 200092, China}
\author{Nathaniel Craig}
\email{ncraig@ucsb.edu}
\affiliation{Department of Physics, University of California, Santa Barbara, CA 93106, USA}
\affiliation{Kavli Institute for Theoretical Physics, Santa Barbara, CA 93106, USA}
\author{Katy Craig}
\email{kcraig@math.ucsb.edu}
\affiliation{Department of Mathematics, University of California, Santa Barbara, CA 93106, USA}
\author{Xinyuan Lin}
\email{xil244@ucsd.edu}
\affiliation{Department of Physics, University of California, San Diego, CA 92093, USA}

\date{\today}

\begin{abstract}
Building upon the success of optimal transport metrics defined for single collinear jets, we develop a multi-scale framework that models entire collider events as distributions on the manifold of their constituent jets, which are themselves distributions on the ground space of the calorimeter. This hierarchical structure of optimal transport effectively captures relevant physics at different scales. We demonstrate the versatility of our method in two event classification tasks, which respectively emphasize intra-jet substructure and inter-jet spatial correlations. Our results highlight the importance of a nested structure of manifolds in the treatment of full collider events, broadening the applicability of optimal transport methods in collider analyses.   
\end{abstract}

\maketitle

\section{Introduction \label{sec:intro}}

Distance-based representations of collider data have been shown to provide powerful inputs to modern machine learning methods (see e.g.~\cite{Larkoski_2020, Guest_2018, albertsson2019machinelearninghighenergy, Radovic2018, Carleo_2019,Bourilkov_2019,Schwartz_2021, karagiorgi2021machinelearningsearchnew, shanahan2022snowmass2021computationalfrontier} for broad overviews). Treating collider events as distributions on the calorimeter naturally suggests the use of optimal transport (OT) distances, which quantify the ``work'' required to rearrange one distribution into another. In recent years, optimal transport methods have opened promising new avenues for analyzing LHC data \cite{Komiske:2019fks, Komiske_2020, Cai_2020, Cai_2022, Komiske:2019jim, Cesarotti:2020hwb, Cesarotti_2021, Cheng_2023, Crispim_Rom_o_2021, ATLAS2023, craig2024exploringoptimaltransporteventlevel, Gouskos_2023, Cesarotti:2024tdh, Gambhir:2024ndc, Ba:2023hix}.\footnote{For a broad overview, see ~\cite{Cai:2023edb}. Other suitable distances between collider events have recently been proposed in \cite{mullin2021doessusyfriendsnew, cai2024phasespacedistancecollider}.  Optimal transport has also been employed to great effect within machine learning architectures for collider physics, see e.g.~\cite{stein2020unsupervised, Kasieczka_2021, Howard_2022, Guglielmo_2021, kansal2022, collins2022exploration}.} 

In the usual application of OT to collider physics, both jets and entire events are considered as transverse momentum distributions in the rapidity-azimuth plane of the calorimeter. Optimal transport distances between these distributions are typically quite effective whenever the underlying task hinges on a single intrinsic scale in the data. However, entire collider events often feature multiple intrinsic scales, such as the scales associated with jet structure and substructure. This raises the natural question: What is the distance between multi-scale collider events?

In this paper, we propose a distance between complete collider events that models them as distributions on the manifold of jets. Inspired by previous work, we endow the manifold of jets with a type of 2-Wasserstein (W$_2$) distance, induced by the  Euclidean ground distance in the rapidity-azimuth plane of the collider. However, unlike previous work, we introduce additional weights which allow us to adjust the relative importance of jet substructure and the locations of the jets within the full collider event. Then, we endow the manifold of events with another 2-Wasserstein distance, induced by the above ground distance on the manifold of jets. In this way, our approach can be thought of as ``W$_2$-on-W$_2$''. This hierarchical structure allows us to effectively distinguish between differences between full collider events, at various scales of the data.

Our paper is organized as follows. \Cref{sec:method} gives an overview of our methodology. First, the usual single-scale optimal transport method is presented, with discussions on suitable representations that account for relevant symmetries acting at the jet and event level. We then introduce the new \emph{multi-scale optimal transport} framework which incorporates both intra-jet substructure and inter-jet spatial correlations for quantifying the dissimilarity between two complete events. \Cref{sec:app} applies the new method to two event classification tasks, i.e., $t \bar t$ vs.~QCD classification and BSM paired dijet vs.~QCD classification, underlining the broad applicability of our method. The paper concludes with~\cref{sec:conclusions}, which highlights several promising directions for future work using the multi-scale OT framework.

\section{Methodology} \label{sec:method}

\subsection{The 2-Wasserstein distance}
\label{subsec:W2_LinW2}

We begin with a brief overview of optimal transport (OT) theory relevant to collider physics, focusing on the $p$-Wasserstein distance and the linearization of the $p=2$ case. 

Consider an event represented by its energy flow, i.e., a discrete distribution $\E = \sum_i E_i \delta_{x_i}$ where the $i$th constituent particle, assumed to be massless, has an energy  $E_i$  and is located at position $x_i$ in some pre-defined {ground space} $\Omega$. Here an ``event'' could denote either a jet or the calorimeter deposit of an entire collision event.\footnote{In the usual application of OT to collider physics, both jets and events are considered as \emph{single-scaled} transverse momentum distributions in the rapidity-azimuth plane, with no distinction  made among possible intrinsic scales in the data. In this subsection, we will refer to both as ``events'', but will distinguish them beginning in \cref{subsec:preprocess_events}.}

Given events $\E$ and $\E'$ with the same total energy, i.e., $\sum_i E_i=\sum_{j'} {E}_{j'}'$, the $p$-Wasserstein distance between them is defined in terms of the  minimization problem
\begin{align} \label{eq:Wp}
    \text{W}_p(\E, \E') & = \min_{\gij\in\Gamma(\E, \E')}\left(\sum_{ij'} \gij \big(d_\text{ground}(x_i, {x}_{j'}')\big)^p \right)^{1/p}, \nonumber \\
    \Gamma(\E, \E') & =\left\{\gij:\gij\geq 0, \sum_{j'}\gij=E_i, 
    \sum_i \gij ={E}_{j'}'\right\}, 
\end{align}
where the \emph{transport plan} $\gij \in \Gamma(\E, \E')$ represents how energy is transferred from particle $i$ in event $\E$ to particle $j'$ in event $\E'$.  The W$_p$ distance measures the difference between two events based on the \emph{minimal effort} required to transfer all the energy from $\E$ to $\E'$, where the \emph{effort} of a transport plan $\gij$ is proportional to $d_\text{ground}(x_i, x_{j'}')$, penalizing the distance over which energy is moved in the ground space. The $g_{ij'}$ that achieves the minimum in~\cref{eq:Wp} is known as the \emph{optimal transport plan}.

It is clear from the definition that the choice of ground metric encodes the essential physics of the problem at hand. In the context of hadron colliders, a natural ground space is the physical rapidity vs.~azimuthal-angle plane of the cylindrical collider, i.e., $\Omega = (y, \phi) \subset \mathbb{R}^2$, with the ground metric being the Euclidean distance $d_\text{ground}(x_i, {x}_{j'}') = \|(y_i,\phi_i) - (y_{j'}', \phi_{j'}') \|$ in this $y-\phi$ plane. Furthermore, the ``energy'' $E_i$ is often replaced by the transverse momentum $p_T$ of the particle. We refer to~\cref{eq:Wp} defined on this ground space as the \emph{single-scale optimal transport} problem.

Different choices of $p$ penalize the ground distance at different length scales.  Two common choices are the W$_1$ metric, also called the Earth Mover's distance ~\cite{Komiske:2019fks, Komiske_2020, Komiske:2019jim, Cesarotti:2020hwb}, and the 2-Wasserstein (W$_2$) distance ~\cite{Cai_2020,Cai_2022,doi:10.1137/21M1400080}. The W$_2$ distance is unique due to its formal Riemannian structure~\cite{lott2007geometric,ambrosio2008gradient,O}, which lends itself conveniently to efficient linearization, as discussed below. Therefore, the default choice in our study is $p=2$.

In general, optimal transport distances are more expensive to compute than classical distances between distributions, such as the $l^2$ norm. For events containing $n$ constituent particles, the Bertsekas’ auction algorithm for the W$_p$ distance demands $\mathcal{O}(n^3)$ operations, while entropic regularization and the Sinkhorn algorithm requires $\mathcal{O}(n^2\log(n))$~\cite{bertsekas1988auction, altschuler2018nearlineartimeapproximationalgorithms, cuturi2013sinkhorndistanceslightspeedcomputation}. In comparison, after binning two distributions, the $l^2$ norm between distributions represented on a grid of size $n$ only requires $\mathcal{O}(n)$ steps. Furthermore, to use OT to compare the mutual differences in a typical collider dataset containing $\mathcal{O}(10^5)$ events,  one would need to calculate $\mathcal{O}(10^{10})$  pairwise OT distances, exacerbating the computational demand.

One way to mitigate this problem is to leverage the Riemannian structure of W$_2$ to \emph{linearize} the distance  \cite{Wang2013-ak,Cai_2020}. In this approach, one projects all events onto the tangent plane of the event manifold at a specific reference event $\mathcal{R} = \sum_i R_i \delta_{y_i}$.\footnote{While the rigorous geometric theory only holds for reference events that don't give mass to small sets,   for simplicity, we explain the approach in the fully discrete case and refer to~\cite{Cai_2020} for more detail.} If $r_{ij}$ denotes the optimal transport plan between a given event $\E$ and the reference $\mathcal{R}$, the projection   is 
\begin{equation}\label{eq:bary}
   \E \mapsto  z_i := \frac{1}{R_i} \sum_j r_{ij}x_j. 
\end{equation}
While the numerical cost of this projection is comparable to the cost of computing a W$_2$ distance, the projection only needs to be computed once per event. Then, all pairwise comparisons between events can be done via cheap $l^2$ norm computations in the tangent plane, i.e., the difference between events $\E$ and $\E'$ is given by
\begin{equation} \label{eq:LinW2}
    \text{LinW$_2$}(\E,\E')=\left(\sum_i R_i||z_i-z_i'||^2\right)^{1/2}.
\end{equation}
While LinW$_2$ itself is not a metric, it converges to a true metric if the reference event satisfies certain properties~\cite[Proposition 1]{Cai_2020}, which are explicitly enforced in the construction of the reference in our study.

\subsection{Representation of jets and events}
\label{subsec:preprocess_events}

In order to use OT to quantify the dissimilarity between energy flows (whether single jets or entire events), one must represent the objects in a manner that takes into account inherent symmetries, so that the OT distance between two distributions that are identical, up to symmetry, is zero. 
  
We consider jets as equivalence classes of distributions $\mathcal{J} = \sum_i p_T^i \delta_{(y_i,\phi_i)}$ in the $y-\phi$ plane, where we identify two jets that are identical up to rotational symmetry.\footnote{For many events, the distribution of jets around their jet axes are correlated, for example by color flow or spin correlations. For simplicity, in the spirit of simplified models, we proceed with our current pre-processing scheme, though such correlation could naturally be included by a mild modification of the ground metric on the jet manifold.} To remove the rotational symmetry on the $y-\phi$ plan, we translate each jet so that its   axis is at $(y, \phi) = (0,0)$, align the principal component of each jet, and then translate each jet back to its original location. The  principal component of each jet~\cite{gurari2011classification, Gallicchio_2013} is given by the principal eigenvector of its moment tensor 
\begin{equation}
    \mathcal{T} = \sum_{i=1}^n p_T^i
    \begin{pmatrix}
         y_i^2 &  y_i \phi_i\\
         y_i \phi_i & \phi_i^2
    \end{pmatrix}. 
\end{equation}
As the principal component is only defined up to a sign, we rotate the jet so that its principal component is along the $\phi$ axis and reflect the jet to ensure that more total $p_T$ resides in the upper part of the $y-\phi$ plane, i.e., $\phi>0$. Note that in the case that $p_T^i = \frac{1}{n}$ for all $i$,  this coincides with the usual principal component analysis (PCA).

In a similar spirit, we consider entire collider events (henceforth ``events'') as equivalence classes of distributions $\E = \sum_s J_s \delta_{\mathcal{J}_s}$ on the manifold of jets, where we identify events that are equal up to symmetry. To build this representation, we must both remove symmetries and translate the complete collision event from a distribution on the plane of the detector into a distribution over its composite jets.

To accomplish this, we first boost an event from the lab frame back to its own center-of-mass (CM) frame, so that the event's CM is at $y=0$. Next, due to the fragmentation of high-energy quarks and gluons, the final-state hadrons of a complete collider event are usually clustered around a few collimated jets. Using the anti-$k_T$ jet clustering algorithm~\cite{Cacciari_2008}, in which a radius $R$ determines the angular extent of a jet, we partition the final-state hadrons of an event $\E$ into distinct jets $\mathcal{J}_s$. The entire collider event is then represented by $\E = \sum_s J_s \delta_{\mathcal{J}_s}$, where there are a variety of choices for how to define the weight $J_s$ of the $s$th jet in the distribution: we devote particular attention to the cases of weighting jets uniformly and in terms of their total $p_T$; see section \ref{sec:ot_Multiple}. Lastly, we translate all jets $\mathcal{J}_s$ simultaneously so that the axis of the jet with the highest $p_T$ is at $\phi=0$. This preprocessing approach guarantees that events differing only by a boost and/or rotation around the beam axis are represented identically.\footnote{Comparing to the jet pre-processing scheme, here we do not rotate an entire event on the $y-\phi$ plane, as such a transformation is not physically meaningful outside the collinear limit. In principle, if the event enjoys a perfect $O(3)$ symmetry, then an extra rotational degree of freedom around the $y,\phi=0$ axis needs to be fixed. One could, for example, further require the axis of the second highest $p_T$ jet to be in the $\phi=0$ plane. However, this symmetry is broken at hadronic colliders as the beam axis (i.e., the $+z$ axis) is unique, due to the fact that the actual four-momenta of the underlying colliding partons are unknown and that a large fraction of final-state particles down the beam line (i.e., close to the $+z$ axis) are not measured by detectors. We therefore do not include this rotational symmetry in our event pre-processing schemes.}

Finally, as described above,  while we typically represent the jets in an event as equivalence classes of distributions, up to rotationally symmetry, in our comparisons of our multi-scale approach with classical single-scale optimal transport, we also consider the case in which the jets are not rotated to align the principle access. We distinguish between these two representations of events in the single-scale optimal transport case as $X_\text{CM+jet}$, when the jets are rotated, and  $X_\text{CM}$, when they are not.

\subsection{Multi-scale Optimal Transport}
\label{sec:ot_Multiple}

\begin{figure}
    \centering
    \includegraphics[width=0.35\textwidth]{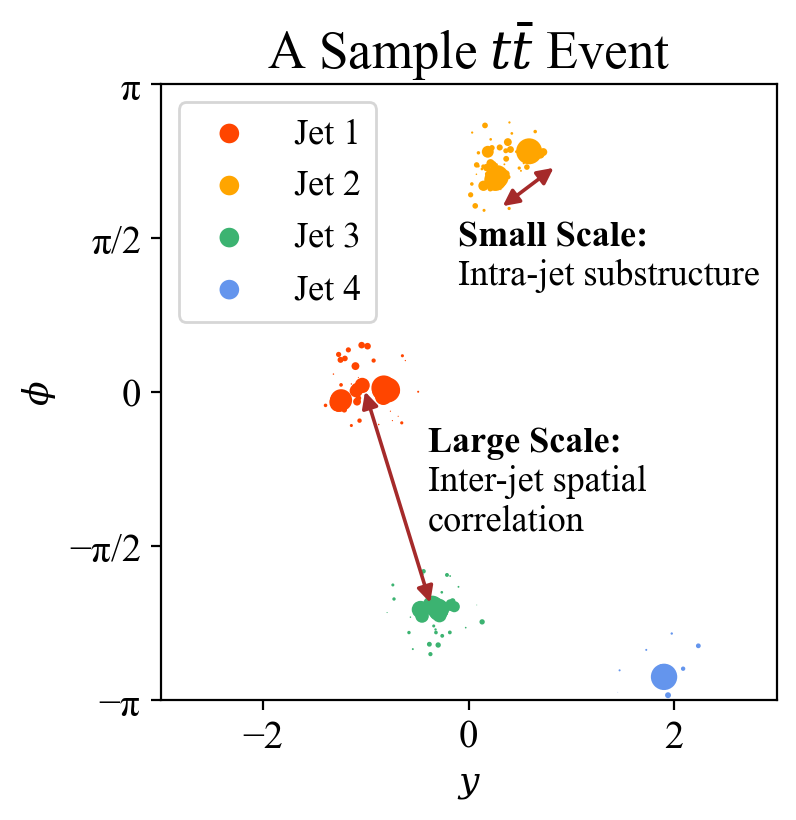}
    \caption{A sample $t \bar{t}$ event composed of four jets (shown in different colors). Constituent particles are displayed on the $y-\phi$ plane, with sizes proportional to their $p_T$. }
    \label{fig:event_multiscale}
\end{figure}

The partition of complete collider events into jets gives  rise to two significant length scales on the $y-\phi$ plane. 
Within a single jet, the relevant length scale is $\mathcal{O}\left(\frac{R}{N}\right)$, where $R \sim \mathcal{O}(1)$ is the jet radius and $N \sim \mathcal{O}(100)$ is particle multiplicity, so that intra-jet substructure is manifested on a length scale of $\mathcal{O}(1/100)$.     
On the other hand, the jets that compose a given event are distributed on the cylindrical calorimeter in a highly non-uniform way, with small clusters separated by large regions mostly devoid of hard particles; see~\cref{fig:event_multiscale}. The separation between jets (red arrow) has a length scale of $\mathcal{O}\left(\frac{C}{M}\right)$, where $C\sim \mathcal{O}(5)$ is the diameter of the detector and $M\sim \mathcal{O}(5)$, leading to an $\mathcal{O}(1)$ length scale for the inter-jet separation. Since this is a hundred times larger than the intra-jet scale, a single-scale W$_2$ distance between entire events as distributions in the $y-\phi$ plane would be dominated by the spatial separation between jets, with little hope to capture the finer distinctions at the substructure level.

Instead, by representing events as distributions on the manifold of jets, i.e., $\E = \sum_{s} J_s \delta_{\mathcal{J}_s}$, we are able to circumvent this issue and include both small-scale substructure within jets and large-scale information about the jets' location on the detector. In particular, the difference between two events $\E$ and $\E'$ can be quantified in terms of a W$_2$ metric, where the ground distance itself is given by a modified W$_2$ distance on the jet manifold. Given $c_{\text{intra}}, c_{\text{inter}} >0$, this is defined as follows:
\begin{align}  \label{eq:W2_event}
    \text{W}_{2,\text{event}}(\E, \E') & = \min_{\pi_{st'} \in\Gamma(\E, \E')}\left(\sum_{st'} \pi_{st'} d^2_{\text{jet}}(\mathcal{J}_s, \mathcal{J}_{t'}') \right)^{1/2},
\end{align}
where
\begin{align}  \label{eq:W2_event}
    d_{ \text{jet}}(\mathcal{J}, \mathcal{J}') &= \Big( \big(c_{\text{intra}} \text{W}_{2,\text{jet}}(\mathcal{J}^*, \mathcal{J'}^*)\big)^2 \nonumber \\
    &+ \big(c_{\text{inter}} \|(\bar{y}, \bar{\phi}) - (\bar{y}', \bar{\phi}') \|\big)^2 \Big)^{1/2}, \nonumber \\
    &  
\end{align}
with $\text{W}_{2,\text{jet}}(\mathcal{J}^*, \mathcal{J'}^*)$ as defined in~\cref{eq:Wp} $(p=2)$ . Here $(\bar{y}, \bar{\phi})$ represents the center of mass of a jet, $\| \cdot \|$ denotes the Euclidean distance on the collider cylinder, and $  \mathcal{J}^*$ and $\mathcal{J'}^*$ represent the translations of jets $\mathcal{J}$ and $\mathcal{J}'$ to have zero center of mass. Due to the fact that two events could, in principle, be composed of a large number of jets, in practice, we use the LinW$_2$ approximation for W$_{2, \text{jet}}$; see~\cref{eq:LinW2}. Furthermore, we use the jet axis from the anti-$k_T$ algorithm as a proxy for the center of mass of the jet. With these choices,~\cref{eq:W2_event} defines the \emph{multi-scale optimal transport} problem.

For any $c_{\text{intra}}, c_{\text{inter}} >0$, $d_{\text{jet}}$ is a metric on the manifold of jets, so W$_{2, \text{event}}$ is a metric on the manifold of complete collider events, where each event is represented as a distribution on the manifold of jets \cite{Ambrosio2013}.
It is a classical result that, when $c_{\text{intra}} = c_{\text{inter}}= c>0$, $d_{\text{jet}}(\mathcal{J},\mathcal{J'}) = c \text{W}_{2,\text{jet}}(\mathcal{J}, \mathcal{J}')$ \cite[Proposition 2.5]{cuesta2003approximation}, so that the ground metric on the jet manifold is exactly the 2-Wasserstein distance between the original untranslated jets. 

We  allow $c_{\text{intra}} \neq c_{\text{inter}}$ in order to capture the effect of the two distinct lengthscales in the composition of an event. A larger ratio of $c_{\text{intra}}/ c_{\text{inter}}$ places a greater emphasis on the substructure (shape) of the jets composing an event, while a smaller ratio of $c_{\text{intra}}/ c_{\text{inter}}$ highlights the relative locations of the jets in the collider plane. As will be seen later, for our specific datasets, $ \|(\bar{y}, \bar{\phi}) - (\bar{y}', \bar{\phi}') \|$ are $\sim O(1)$ whereas W$_{2,\text{jet}}(\mathcal{J}^*, \mathcal{J'}^*)$ are $\sim O(0.1)$. Thus, the choice $(c_{\text{intra}}, c_{\text{inter}}) = (10,1)$ roughly gives equal weighting on both terms. 

As described in~\cref{subsec:preprocess_events}, when representing an event as a distribution on the manifold of jets, $\E = \sum_s J_s \delta_{\mathcal{J}_s}$, there are several potential choices one could make for the weights $J_s$ of each jet $\mathcal{J}_s$. A first choice would be to define
\begin{align*}
J_s = \begin{cases} 1 &\text{ if } \mathcal{J}_s \text{ is the hardest jet,}  \\ 0 &\text{ otherwise,} \end{cases}
 \end{align*}
 where by \emph{hardest jet}, we indicate the jet with the largest total $p_T$. In this case, the 2-Wasserstein distance between events simplifies to 
\begin{align} \label{eq:distEE'_hardestjet}
  \text{W}_{2,\text{event}}(\E, \E') = \Big( \big(c_{\text{intra}}  \text{W}_{2,\text{jet}}(\mathcal{J}_1^*, \mathcal{J}_{1'}'^*)\big)^2 \nonumber \\
  + \big(c_{\text{inter}}  \|(\bar{y}_1, \bar{\phi}_1) - (\bar{y}_{1'}', \bar{\phi}_{1'}') \| \big)^2 \Big)^{1/2}, 
\end{align}
where $\mathcal{J}_1$ is the jet with the highest $p_T$ in event $\E$ and $\mathcal{J}_{1'}'$ for event $\E'$. While this choice of weighting is sensitive to both the substructure difference between the two hardest jets and their spatial separation on the $y-\phi$ plane, it does not  take into account any of the features of softer jets. This simple method will be referred to as ``Multi-scale OT: Jet 1''.

More sophisticated choices of weighting are to either choose $J_s$ uniformly, letting $J_s \equiv 1/M$, where $M$ is the number of jets in event $\E$, or to choose $J_s$ proportional to the transverse momentum of the $s$th jet, i.e.,
\begin{align*}
J_s = \frac{p_T^{s \text{th jet}}}{\sum_{s'} p_T^{s' \text{th jet}} }.
\end{align*}
For this second approach, jets with higher $p_T$ are more influential in determining the overall OT distance between events. In what follows we will refer to these approaches as `Multi-scale OT: uniform weighted OT'' and   ``Multi-scale OT: $p_T$ weighted OT,'' respectively.

\subsection{A toy example}
\label{subsec:toy}

\begin{figure*}[ht]
    \begin{subfigure}{0.75\textwidth}
    \includegraphics[width=\textwidth]{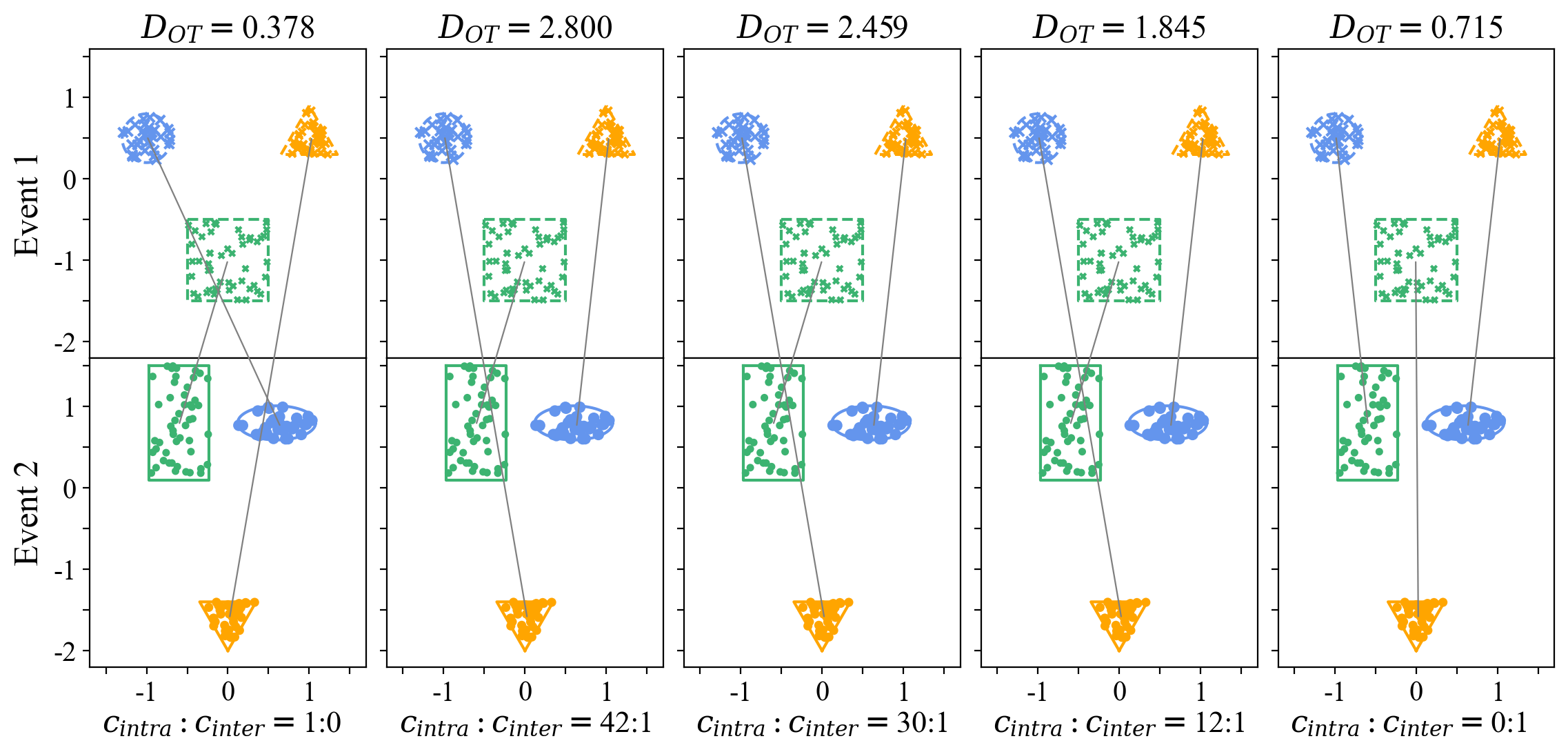}
    \caption{} \label{fig:toymodel_c}
    \end{subfigure}
    \begin{subfigure}{0.23\textwidth}
    \includegraphics[width=\textwidth]{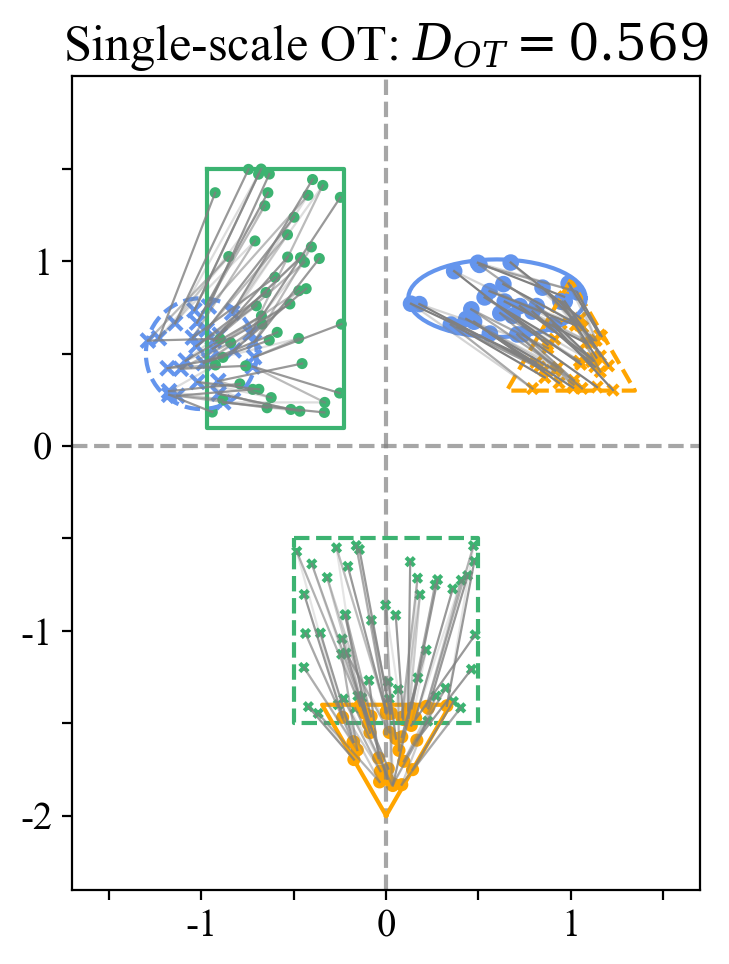}
    \vspace*{2mm}\caption{} \label{fig:toymodel_b}
    \end{subfigure}
    \caption{(a) A series of multi-scale OT distances between the two events under different $(c_{\text{intra}}, c_{\text{inter}})$ pairs, where only the event-level transport plans are shown, denoting the matching among jets as a whole; (b) Single-scale OT between the two events, showing the transport plan among the particles between the two events (represented by the gray lines).} 
    \label{fig:toymodel}
\end{figure*}

We close our discussion of the method with an illustration of its performance on a toy example of two artificial ``events''. Here each event is composed of three jets, and we want to quantify the dissimilarity of the events in terms of both small-scale (shape of jets) and large-scale (physical location of jets) features. All particles that make up each jet are equally weighted, and the event is considered as a uniformly weighted distribution on its constituent jets. We choose a simple Euclidean $x-y$ plane as the ground space. We compare the performance of a classical single-scale W$_2$ distance between the distributions with our new multi-scale OT distance. In both cases, the distance is determined by the corresponding optimal transport plan. Consequently,  we illustrate the differences by comparing the structure of the optimal transport plans in each case. Further details of the setup are given in~\cref{app:details_toy}.    

We first compute the single-scale W$_2$ distance between $\E$ and $\E'$, using~\cref{eq:Wp} with $p=2$. \Cref{fig:toymodel_b} shows the optimal transport plan among all particles, which only considers the vicinity of the particles on the $y-\phi$ plane.  To also capture the difference in the shape of the jets, we need the multi-scale OT distance. In~\cref{fig:toymodel_c}, we consider five pairs of coefficients,
$$(c_{\text{intra}}, c_{\text{inter}})=(1,0),(42,1),(30,1),(12,1),(0,1),$$
which are chosen to illustrate the role of small scale vs.~large scale effects. As expected, when $c_{\text{intra}}/ c_{\text{inter}}$ is small, the event level optimal transport plan prefers to match the jets between the events based on the location of the jet in physical space. Conversely, when $c_{\text{intra}}/ c_{\text{inter}}$ is large, the event level optimal transport plan prefers to match the jets between the events based on the similarity of their shape.

\section{Application to Event Classification} \label{sec:app}

We now apply both single-scale and multi-scale OT frameworks to two different event classification tasks to illustrate the utility of the latter. The first task aims to distinguish boosted $t \bar{t}$ events from QCD events, while the second task aims to distinguish a pair-produced BSM particle decaying into dijets from QCD. At a typical jet radius of $R =0.5$, the two tasks respectively benefit from intra- and inter-jet information. 

Although we use $R = 0.5$ jets for our primary benchmark, it is also illuminating to see how intra- and inter-jet contributions change as the jet radius is varied. As $R$ decreases, inter-jet information becomes more important as decays of parent particles are increasingly resolved and the radiation pattern within each jet exhibits less structure. Conversely, intra-jet information becomes more important as $R$ increases and the decay products of parent particles are clustered together. 

Once a distance between events is obtained using either of the two OT frameworks, a classifier is needed to take the distance as input and output the class label for each event in a supervised manner. Given that the focus of our study is on the role of the metrics between events, not on the classifier, we simply use the k-nearest neighbor (kNN) algorithm \cite{Cover1967NearestNP}, as it depends solely on the notion of pairwise distances. kNN classifies a new data point based on the majority of its $k$ nearest neighbors, with neighbors defined by the user-specified distance. Other more sophisticated classifiers, including neural networks, can be coupled to both single-scale and multi-scale OT frameworks; we leave this for future study.

Our dataset for each classification task always consists of 5000 background events and 5000 signal events, further split into $50\%$ for training the classifier, $25\%$ for validation, and $25\%$ for testing. The optimal value of the model hyper-parameter $k$ is selected using the validation dataset, with $k$ ranging from 10 to 200 with an increment of 10. The performance of kNN is evaluated on the held-out test set, which outputs the final AUC values.

For comparison with non-OT methods, we choose the classical event shape observable \emph{transverse thrust}~\cite{Bertram:2002sv, Nagy_2003, Banfi_2004}, which is widely-used in $pp$ collisions and served as a benchmark for OT-based event isotropy studies~\cite{Cesarotti:2020hwb}. The transverse thrust is defined as
\begin{equation}
    T_\perp = \max_{\hat{n}_\perp}\frac{\sum_i|\vec p_{T,i} \cdot \hat{n}_\perp|}{\sum_i |\vec p_{T,i}|}
\end{equation}
where $\vec p_{T,i}$ is the transverse momentum of the $i$th particle in the event, and $\hat{n}_\perp$ is a unit vector on the transverse plane. $T_\perp$ ranges from $\frac{2}{\pi}$ for a fully isotropic event, to $1$ for a completely collinear event. For convenience and following the convention in~\cite{Cesarotti:2020hwb}, we use a rescaled version of $T_\perp$ so that the range becomes $[0,1]$. That is, each event corresponds to a single scalar 
\begin{equation}
    \Tilde{T}_\perp = \frac{\pi}{\pi - 2 }\left(T_\perp - \frac{2}{\pi}\right)
\end{equation}
We apply a cut on $\Tilde{T}_\perp$ to separate signal events from background events. A sweep over the cut value then outputs the AUC value for the classification using this $\Tilde{T}_\perp$ observable, which may be compared with the two OT methods.

\subsection{$t \bar{t}$ \textit{vs.}~QCD Event Classification}
\label{subsec:QCDtt}

All events are generated in \texttt{MadGraph 2.9.16} \cite{Alwall_2014} using $pp$ collisions with a center of mass energy $\sqrt{s}=13$ TeV. For QCD events, we consider $pp \to jjjj$, where a jet $j$ is originated by either a gluon or $u, d, c, s$ quarks (and the corresponding anti-quarks). For $t\bar{t}$ pair production, we consider $pp \to t\bar{t} \to W^+ b W^- \bar{b} \to (jj)(jj)bb$ events, where the boosted $W$ boson decays hadronically. To increase event yield, a minimum $p_T$ cut of $150$ GeV is applied to each $j$ product of the QCD events, whereas a $p_T>450$ GeV cut is applied to the top quarks themselves.   
Hadronization and showering are done in \texttt{Pythia 8.310} \cite{bierlich2022comprehensiveguidephysicsusage} with default parameters. Events are then clustered into jets with \texttt{FastJet 3.4} \cite{Cacciari_2012}, using the anti-$k_T$ algorithm~\cite{Cacciari_2008} with a jet radius of $R=0.5$. Only events with at least four jets are retained. A rapidity cutoff of $y \leq 4$ is applied to each jet. The two highest $p_T$ jets (labeled \texttt{Jet 1} and \texttt{Jet 2}) are required to have $p_T>300$ GeV, whereas the next two jets (labeled \texttt{Jet 3} and \texttt{Jet 4}) have a threshold of $p_T>150$ GeV. In cases where more than four jets are identified in an event, we only keep the first four jets with the largest $p_T$ and discard the rest. These four jets define an event in our dataset. 

In order to study the effect of varying jet radius, we further create two more datasets based on the above dataset. For each event in the above $R=0.5$ dataset containing exactly four jets, we recluster all its constituent particles using the anti-$k_T$ algorithm with two new radii, respectively $R=0.3$ and $R=1.0$. We emphasize that this procedure {\it only} includes particles that had been clustered in the leading four $R = 0.5$ jets of each original event.\footnote{The jet mass distribution of $R = 0.5$ jets and the impact of reclustering on jet mass and multiplicity distributions is illustrated in~\cref{app:details_ttqcd}.} Although artificial, this provides the clearest illustration of the interplay between the jet radius $R$ and the contributions from intra- and inter-jet distributions. 

\begin{figure}[h]
    \centering
    \includegraphics[width=0.5\textwidth]{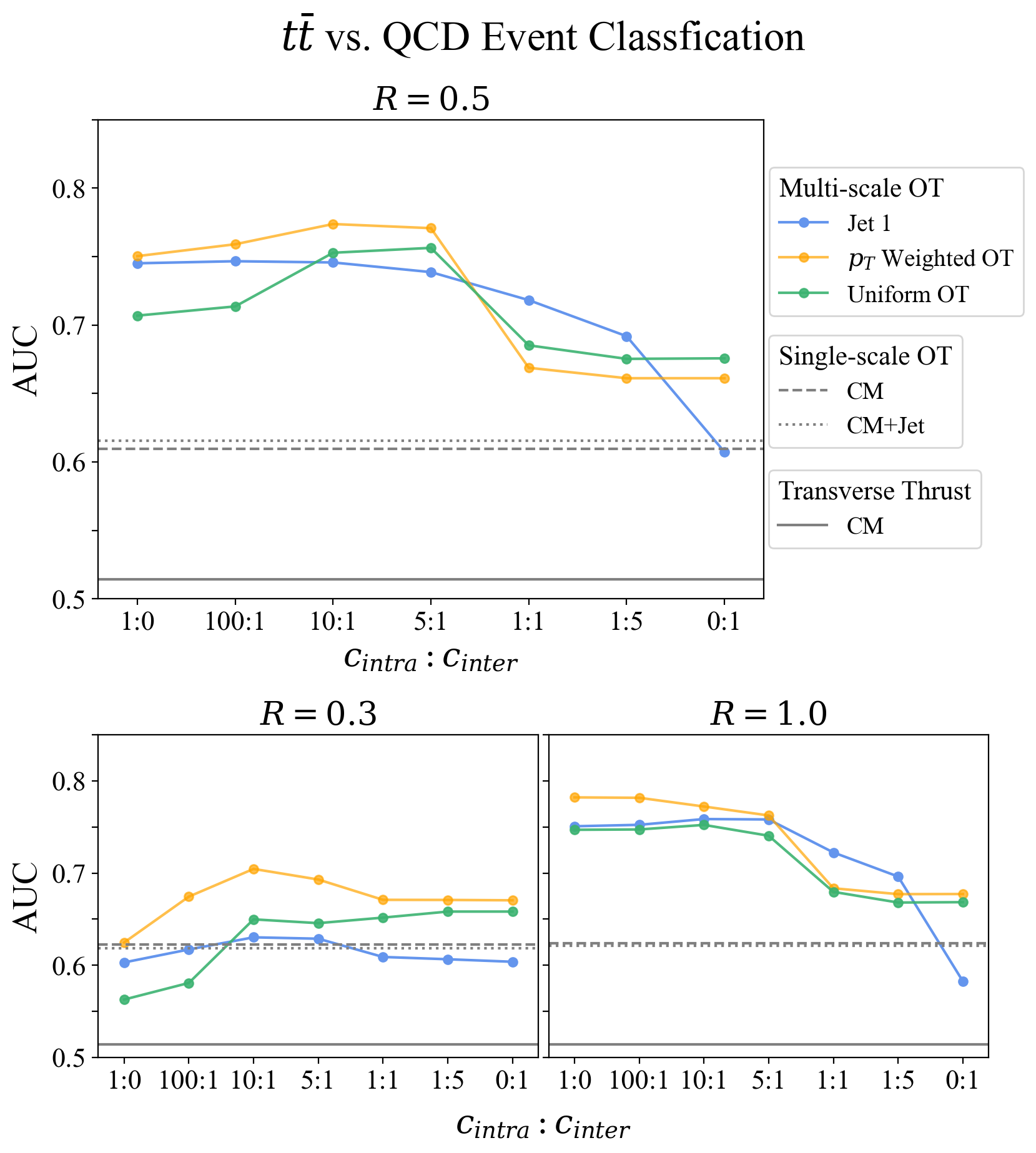}
    \caption{AUC scores for the $t \bar{t}$ \textit{vs.}~QCD event classification task, when events are clustered into jets with radius $R=0.5$ (\textit{upper}), $0.3$ (\textit{lower left}), and $1.0$ (\textit{lower right}). The colored curves denote the respective results for the three multi-scale OT methods, i.e., ``$p_T$ weighted OT'', ``uniform weighted OT'', and ``Jet 1'', as we vary the values for the coefficient pair $(c_\text{intra}, c_\text{inter})$ (expressed in terms of ``$c_\text{intra}:c_\text{inter}$'' on the $x$-axis). The performance of the single-scale OT framework are shown in grey dashed lines, with the two different pre-processing schemes, i.e., ``CM'' and ``CM+Jet''. Also included as benchmark is the performance of the transverse thrust (grey solid line), where events are pre-processed according to the ``CM'' scheme.}
    \label{fig:auc_QCDtt}
\end{figure}

\Cref{fig:auc_QCDtt} shows the classification performance (measured in AUC) on the test set for both the single-scale and multi-scale OT frameworks coupled to a kNN classifier, where jets are clustered with different radii $R=0.3, 0.5, 1.0$. Also displayed for comparison is the performance of transverse thrust, which stays the same for all three datasets as the underlying events are the same.

When using the single-scale OT method, events are pre-processed according to either the ``CM'' or the ``CM+jet'' schemes introduced in~\cref{subsec:preprocess_events}. As can be seen, the two pre-processing schemes make little difference to the classification performance. This suggests that the additional jet-level pre-processing step generates no meaningful improvement under the single-scale OT framework. Such behavior is to be expected, as the existence of multiple scales intrinsic to a full collision event is being ignored.  

The multi-scale OT framework shows improved performance in almost all cases considered. One interesting coefficient pair is $(c_{\text{intra}}, c_{\text{inter}}) = (1, 0)$, where no inter-jet spatial correlation is used and the information given to both single-scale and multi-scale OT frameworks is the same. In spite of the same inputs, for $R=0.5$ and $1.0$, the multi-scale OT method still outputs an AUC about $25\%$ higher than that of the two single-scale benchmarks. Such a large improvement underlines the importance of the ability to highlight substructure variation via the ground metric of~\cref{eq:W2_event}. 

For more general choices of $(c_{\text{intra}}, c_{\text{inter}})$, all three combination methods under the multi-scale framework (i.e., ``$p_T$ weighted OT'', ``uniform weighted OT'', and ``Jet 1'') give relatively comparable performance across different $R$ values, with slight preference to ``$p_T$ weighted OT''. That ``uniform weighted OT'' matches the performance of ``$p_T$ weighted OT'' is to be expected, as all included jets in a given event have more or less similar $p_T$, especially for larger $R$. On the other hand, the ``Jet 1'' method performs well overall in part because of the nature of this particular classification task, as most of the hardest jets are $W$ or $t$ jets for $R=0.5$ and $1.0$ whose substructure alone contains enough information to identify signal events from backgrounds. In~\cref{subsec:QCDBSM}, we will see that this is sensitive to the classification task.

The three multi-scale methods perform best when more weight is put on intra-jet substructure, which contains key discriminative information for this task. Indeed, for $R=1.0$, the best AUC of 0.782 is obtained with the coefficient pair $(c_{\text{intra}}, c_{\text{inter}}) = (1, 0)$, and the performance drops when more emphasis is given to inter-jet spatial correlation. For smaller $R=0.3$ or $0.5$, the highest AUC is achieved with the ``$p_T$ weighted OT'' method around $(c_{\text{intra}}, c_{\text{inter}}) = (10, 1)$, which is roughly where intra-jet and inter-jet parts contribute equally. Particularly for $R=0.3$, the decay products are expected to become more resolved and the corresponding jets have less substructure with which to distinguish them from QCD jets. In this case, inter-jet spatial correlation becomes more important, as verified by the differing trend of the curves with respect to the $(c_{\text{intra}}, c_{\text{inter}})$ ratio, comparing to larger $R$ values.

Additionally, for $R=0.5$ and $1.0$, the performance of the ``$p_T$ weighted OT'' and ``uniform weighted OT'' methods decline sharply around $(c_{\text{intra}}, c_{\text{inter}}) = (1, 1)$, whereas for the ``Jet 1'' method, the AUC displays a noticeable drop only when there is   no contribution from intra-jet substructure. In these cases, when the  jet radii are larger, the highest $p_T$ jet is more likely to be a complete boosted jet, i.e., a top jet, and its substructure carries more discriminative information. Since the event preprocessing procedure aligns the axis of its hardest jet, the  inter-jet spatial correlation  contributes less to the   event W$_2$ distance for the ``Jet 1'' method than the other two multi-scale methods.
Consequently, while the other multi-scale schemes deteriorate as too much emphasis is placed on the (less informative) inter-jet differences,   “Jet 1” doesn’t deteriorate until ratio (0,1), when events are compared exclusively based on inter-jet location differences (ignoring substructure). Furthermore, at ratio (0,1), “Jet 1” is worse than other multi-scale approaches, since, in this case, the method only classifies based on the inter-jet location differences of a single jet in each event, as compared to the other multi-scale methods which consider multiple jets in each event.The same phenomenon also occurs in the BSM \textit{vs.}~QCD event classification with $R=1.0$ in~\cref{subsec:QCDBSM}, for similar reasons.

Finally, even when $(c_{\text{intra}}, c_{\text{inter}}) = (0, 1)$, multi-scale OT (with the exception of the ``Jet 1'' method as explained above) still gives noticeably better performance than single-scale OT, again affirming the importance of a multi-scale treatment.

\subsection{BSM \textit{vs.}~QCD Event Classification}
\label{subsec:QCDBSM}

The second classification task involves discrimination between a paired dijet signal from BSM physics and the corresponding four-jet Standard Model QCD background. Whereas jet substructure played a significant role in the first classification task, here the jet substructure plays a less decisive role compared to the jet spatial distributions themselves.

We consider a new BSM scalar particle $\phi$, which carries electromagnetic charge $+ \frac{1}{3}$ and has mass $m_\phi = 100$ GeV with width $\Gamma_\phi = 2$ GeV. It is pair produced via $pp \to \phi \bar{\phi}$ and subsequently decays into two QCD jets, with equal coupling to all quark pairs that conserve charge. The Feynman rules for the new particle are computed using \texttt{FeynRules}\cite{Alwall_2014}, and passed to \texttt{MadGraph} for the computation of the matrix elements. Both the BSM signal events and QCD background events involve a $4j$ final state, and a minimum jet $p_T$ of $150$ GeV is imposed at simulation level. Hadronization and showering are again conducted with \texttt{Pythia} and jet clustering is performed in \texttt{FastJet} using the anti-$k_T$ algorithm with $R=0.5$. A rapidity cutoff of $y \leq 4$ and a $p_T$ threshold of $p_T \geq 150$ GeV for each jet are applied. As before, we only select events with at least four jets and retain only the four largest-$p_T$ jets for each chosen event. As with the first classification task, we create two additional datasets by reclustering the particles in the $R = 0.5$ dataset using $R = 0.3$ and $R=1.0$. The jet mass distribution of $R = 0.5$ jets and the impact of reclustering on jet mass and multiplicity distributions is illustrated in~\cref{app:details_bsmqcd}.

\begin{figure}[h]
    \centering
    \includegraphics[width=0.5\textwidth]{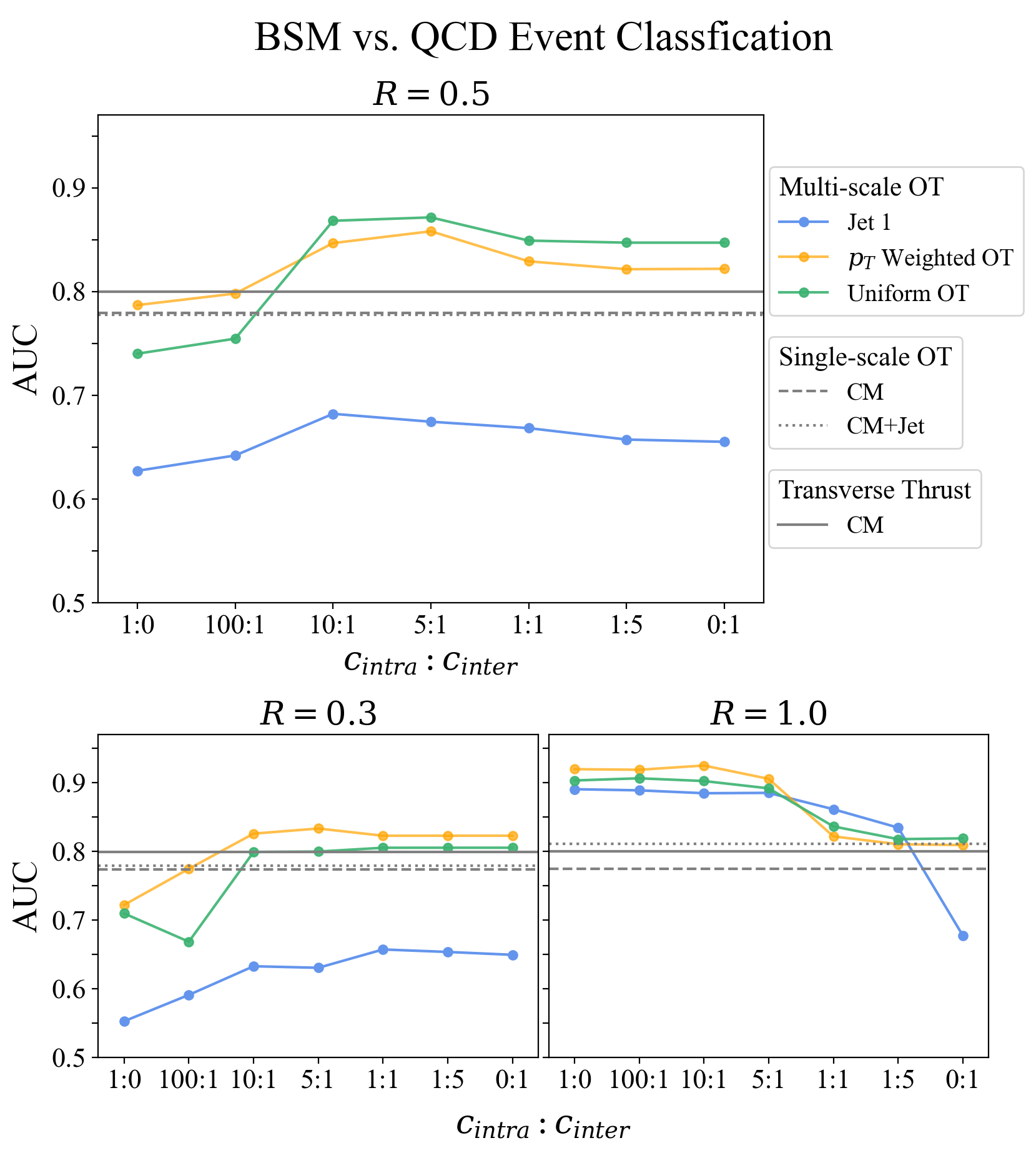}
    \caption{AUC scores for the BSM \textit{vs.}~QCD event classification task, when events are clustered into jets with radius $R=0.5$ (\textit{upper}), $0.3$ (\textit{lower left}), and $1.0$ (\textit{lower right}). The colored curves denote the respective results of the three multi-scale OT methods, i.e., ``$p_T$ weighted OT'', ``uniform weighted OT'', and ``Jet 1'', as we vary the values of the coefficient pair $(c_\text{intra}, c_\text{inter})$ (expressed in terms of ``$c_\text{intra}:c_\text{inter}$'' on the $x$-axis). The performance of the single-scale OT framework are shown in grey dashed lines, with the two different pre-processing schemes, i.e., ``CM'' and ``CM+Jet''. Also included as benchmark is the performance of the transverse thrust (grey solid line), where events are pre-processed according to the ``CM'' scheme.}
    \label{fig:auc_QCDBSM}
\end{figure}

\Cref{fig:auc_QCDBSM} shows the performance of transverse thrust, single-scale OT, and multi-scale OT methods on the BSM \textit{vs.}~QCD classification task. Here, the benchmark observable transverse thrust already performs well, with AUC around $0.8$ for all three $R$ choices. It is on the same par with (and even slightly better than) the single-scale OT framework. 

The $p_T$-weighted and uniform-weighted multi-scale OT methods exhibit the best performance for optimal values of $c_\text{intra}:c_\text{inter}$, and the interplay between jet radius and $c_\text{intra}:c_\text{inter}$ is as expected for these methods. For $R = 0.5$ the decay products of each $\phi$ are typically resolved, favoring larger $c_{\rm inter};$ this persists for smaller values of $R$. Conversely, by $R = 1.0$ the decay products of each $\phi$ are typically clustered into fat jets, favoring larger $c_{\text{intra}}$. That being said, the optimal performance for $R=0.5$ is achieved around $(c_{\text{intra}}, c_{\text{inter}}) = (10, 1) \text{ and } (5, 1)$, where both inter-jet and intra-jet contributions are comparable; additional information is captured by the intra-jet component, given that the decay products from some $\phi$ decays are clustered together.   

In contrast, the ``Multi-scale OT: Jet 1'' method performs poorly at and below $R = 0.5$. This is to be expected, as at $R = 0.5$ the hardest jets in both signal and background events originate from individual quarks or gluons. For $R = 1.0,$ however, the decay products of the $\phi$ are typically collected into fat jets, and the ``Jet 1'' method becomes comparable to the other two multi-scale OT methods.

\section{Conclusions and Future Directions} \label{sec:conclusions}

Optimal transport has proven to provide useful metrics between individual jets at hadron colliders~\cite{Komiske:2019fks, Komiske_2020, Cai_2020, Cai_2022, Komiske:2019jim, Cesarotti:2020hwb, Cesarotti_2021,Crispim_Rom_o_2021, ATLAS2023, craig2024exploringoptimaltransporteventlevel, Gouskos_2023}. In this work, we have introduced a new multi-scale OT framework that extends the original single-scale optimal transport methods to complete collision events, effectively leveraging information encoded in both intra-jet and inter-jet distributions. We have achieved this by first representing jets as distributions on the calorimeter, and entire events as distributions on the manifold of jets.  

We demonstrated the versatility of the multi-scale OT distance by considering two event classification tasks of different natures. For almost all cases considered, the multi-scale OT framework is consistently better than the original single-scale OT distance and the non-OT benchmark of transverse thrust. For standard $R = 0.5$ jets, the best performance is obtained by either $p_T$-weighted or uniformly-weighted multi-scale methods, with numerically comparable contributions from inter-jet and intra-jet structure. The broad efficacy of these methods suggests they may be useful in general search and classification tasks at colliders.    

There are various avenues to further development. For our current work, each collision event is first normalized to have the same total ``mass'' of one unit, before computing the balanced 2-Wasserstein distance with~\cref{eq:W2_event}. Such normalization ignores the potentially relevant information stored in the total ``mass'' difference between two events, with ``mass'' being either the total scalar $p_T$ of the event in the ``$p_T$ weighted OT'' scheme, or the jet multiplicity in the ``uniform weighted OT'' scheme. Analogous to~\cite{doi:10.1137/21M1400080, cai2025multi}, which used the Hellinger-Kantorovich extension of the 2-Wasserstein distance to compare individual jets with different total $p_T$, the Hellinger-Kantorovich approach could also be used in the present setting  to capture the additional  information  of the total ``mass'' difference between events, and this  may  improve the performance of our current multi-scale OT framework.     

In collider experiments, one typically has access to more information than just calorimetry, such as particle charge and tracks. Such additional information can in principle also be incorporated into the definition of an optimal transport distance, and may significantly improve the sensitivity of the distance to the underlying hard process. Another future direction would be to combine our novel notion of distance between complete collider events with the optimal transport approach for background modeling developed by Manole, et. al. \cite{manole2024backgroundmodelingdoublehiggs}, which is similar in spirit to our approach, due to its nested use of  optimal transport.

More generally, the multi-scale OT framework developed here can be applied to many scenarios within and beyond collider physics. Any data that features multiple length scales and can be described in terms of distributions on similarly nested manifolds could benefit from a comparable approach.

\begin{acknowledgments}
The authors thank the referee for their valuable suggestions. TC would like to further thank Rikab Gambhir for suggesting the inclusion of missing references. The work of KC has been supported by the National Science Foundation under Grant No.~NSF DMS-2145900. The work of TC and NC was supported in part by the U.S.\ Department of Energy under the grant DE-SC0011702 and performed in part at the Kavli Institute for Theoretical Physics, supported by the National Science Foundation under Grant No.~NSF PHY-1748958. TC was additionally supported by the U.S. Department of Energy Awards No.~DE-FOA-0002705, KA/OR55/22 (AIHEP) and No.~DE-AC02-76SF00515. TC also gratefully acknowledges the support of the SCale Family Fund during the career transition period. 
\end{acknowledgments}

\appendix

\section{Toy Example} \label{app:details_toy}

\begin{figure}
    \centering
    \includegraphics[width=0.4\textwidth]{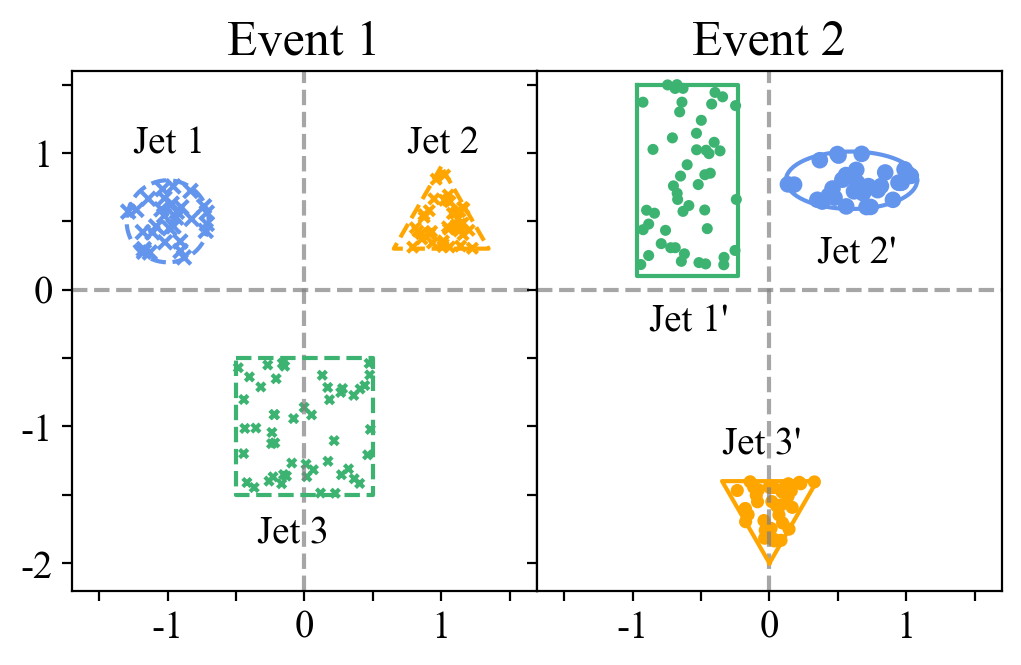}
    \caption{Two artificially generated events. Each event contains three jets, highlighted in different colors and with their overall shape delineated for illustrative purposes.}
    \label{fig:enter-label}
\end{figure}

Here we give more details on the two artificial events in the toy example in~\cref{subsec:toy}. Both events consist of a circular ``jet'', a triangular ``jet'', and a rectangular ``jet'', mimicking the morphological difference of real jets; see~\cref{fig:enter-label}. The total ``mass'' of each of the three jets is set to be $\frac{1}{3}$, giving unit mass to both events. The particles within each jet all have the same mass $\frac{1}{3n}$, where $n$ is the number of particles inside each jet.

In particular, the particles in the first event $\E$ are:
\begin{itemize}
    \itemsep-0.3em 
    \item Jet 1 (circular): 30 points uniformly sampled within a circle centered at $(-1, 0.5)$ with a radius of $0.3$;
    \item Jet 2 (triangular): 40 points uniformly sampled within an equilateral triangle centered at $(1,0.5)$ with a height of $0.6$;
    \item Jet 3 (rectangular): 50 points uniformly sampled within a square centered at $(0,-1)$ with a side of $0.6$.
\end{itemize}
The locations of the three jets are chosen so that the center of mass of the whole event is approximately at $(0,0)$. The second event $\E'$ is generated with slight modifications to $\E$:
\begin{itemize}
    \itemsep-0.3em
    \item Jet $1'$ (rectangular): 50 points uniformly sampled within a rectangle centered at $(-0.6, -0.8)$ with sides being $1.0$ and $0.4$;
    \item Jet $2'$ (circular): 30 points uniformly sampled within an ellipse centered at $(0.6, 0.8)$ with axes being $a=0.48$ and $b=0.21$;
    \item Jet $3'$ (triangular): 40 points uniformly sampled within an upside down equilateral triangle centered at $(0, -1.6)$ with height again being $0.6$.
\end{itemize}
These numerical choices for the jet locations are made so that the center of mass of the complete event stays approximately the same for both events.

\section{$t \bar t$ vs. QCD Events} \label{app:details_ttqcd}

In this and the next appendix, we provide supplemental information about the events used in the $t \bar{t} $ vs.~QCD and BSM vs.~QCD classification tasks, respectively. 

\begin{figure}[h]
    \centering
    \includegraphics[width=0.495\textwidth]{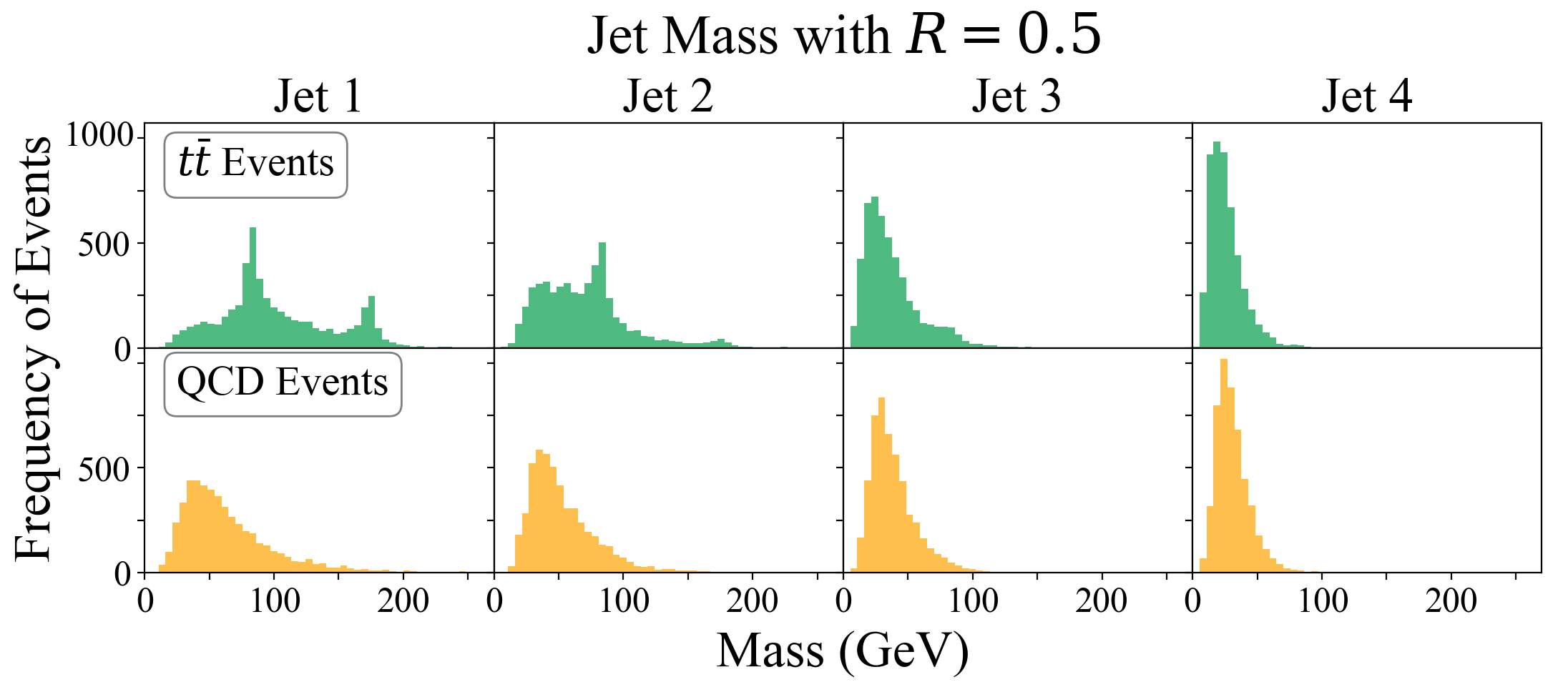}
    \caption{Mass of the first four highest $p_T$ jets (clustered with $R=0.5$) in the signal $t \bar{t}$ events (\textit{top}) and the background QCD events (\textit{bottom}).}
    \label{fig:tQCD_jetmass_r0.5}
\end{figure}

The jet mass distributions of the four $p_T$-ordered $R= 0.5$ jets passing selection cuts in both the $t \bar t$ and QCD event samples are shown in~\cref{fig:tQCD_jetmass_r0.5}. A noticeable fraction of the highest-$p_T$ \texttt{Jet 1} in the signal events are top jets, peaking around a mass of $\sim 170$ GeV, together with a majority of $W$ jets with a mass peak around $80$ GeV. Many of \texttt{Jet 2} are also $W$ jets, while both \texttt{Jet 3} and \texttt{Jet 4} are mostly QCD jets. We therefore expect intra-jet substructure to play a significant role in the classification task.

\begin{figure}[h]
    \begin{subfigure}{0.232\textwidth}
    \includegraphics[width=\textwidth]{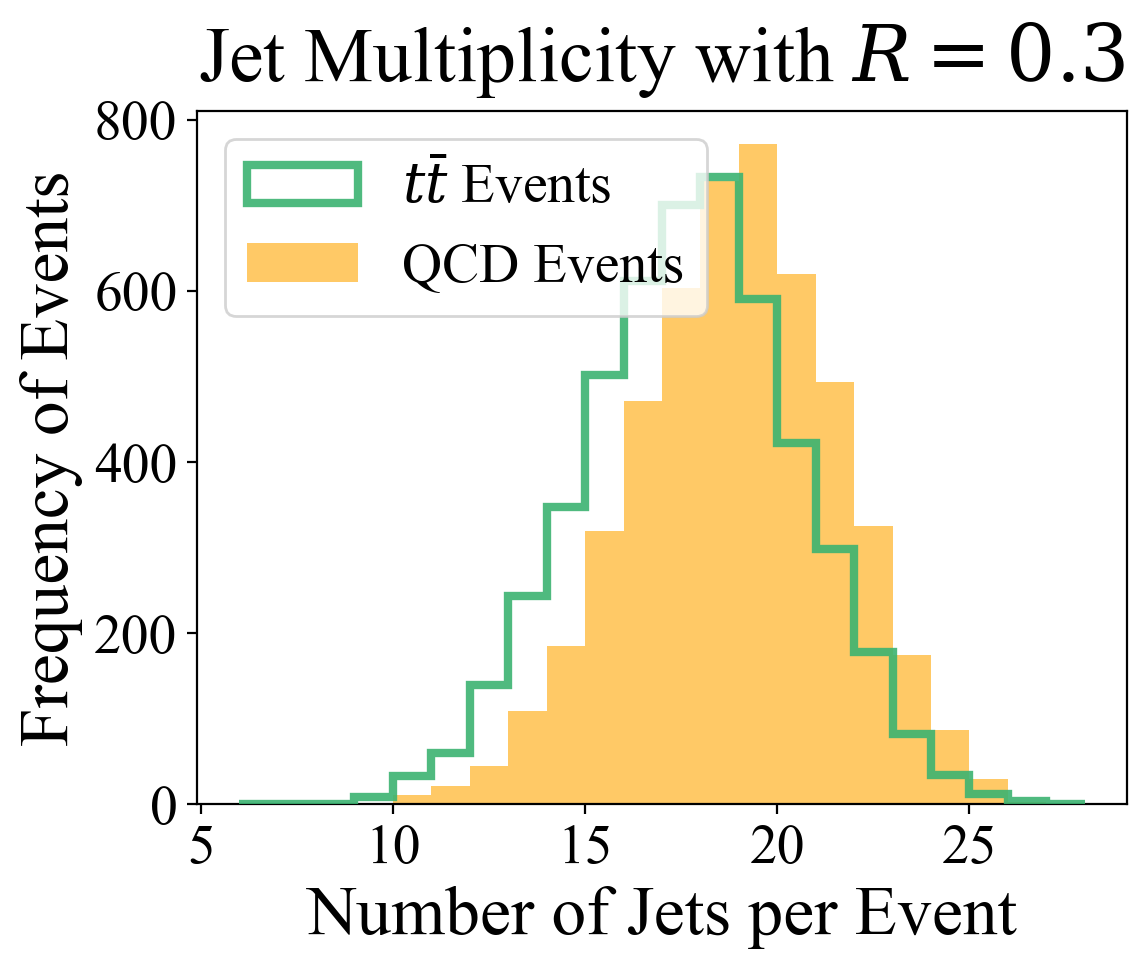}
    \caption{} \label{fig:tQCD_diffr_a}
    \end{subfigure}
    \hspace*{\fill} 
    \begin{subfigure}{0.238\textwidth}
    \includegraphics[width=\textwidth]{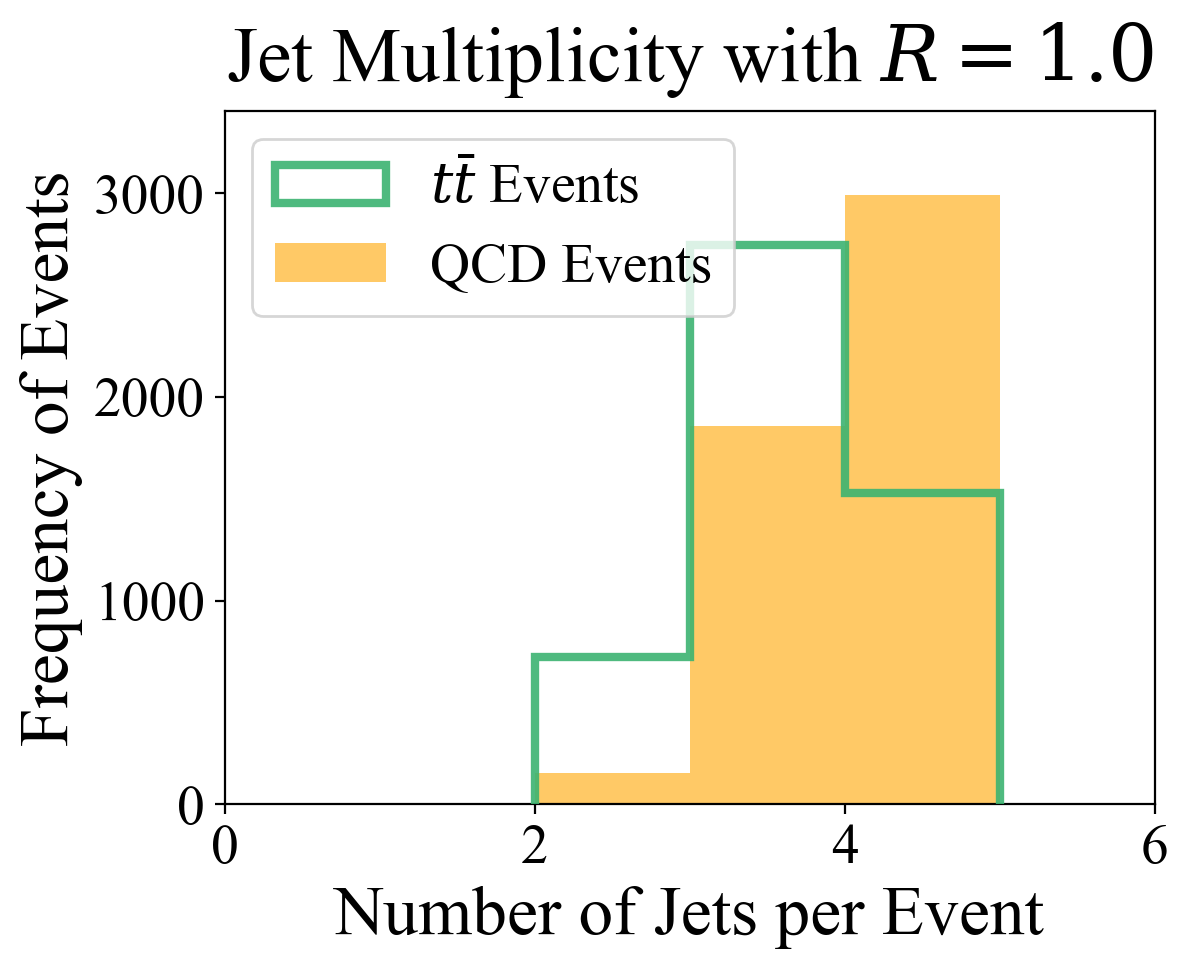}
    \caption{} \label{fig:tQCD_diffr_b}
    \end{subfigure}\\
    \begin{subfigure}{0.5\textwidth}
    \includegraphics[width=\textwidth]{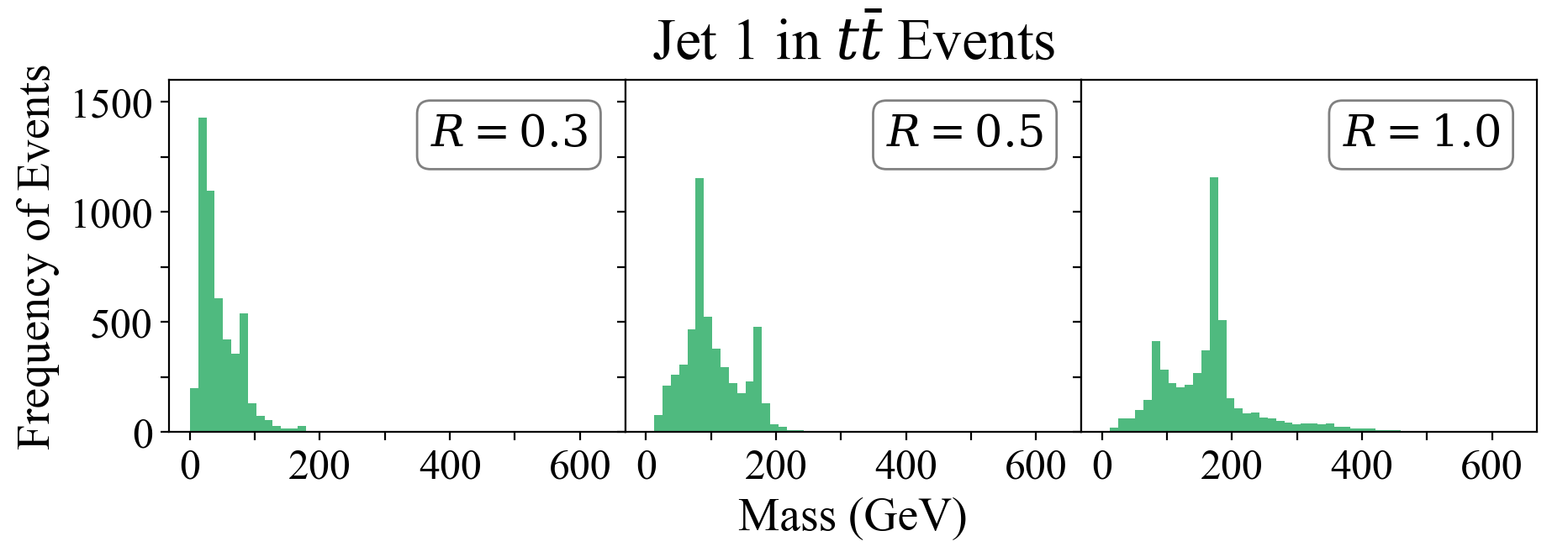}
    \caption{} \label{fig:tQCD_diffr_c}
    \end{subfigure}
    \caption{Jet multiplicity in each event, where jets are clustered with a radius of (a) $R=0.3$, or (b) $R=1.0$. Note that the label on the left tick of the $x$-axis shows the number of jets in each bin, i.e., for example, in (b) roughly 700 $t \bar{t}$ events have 2 jets with $R=1.0$. (c) Jet mass for the first largest $p_T$ jet in the signal $t \bar{t}$ events (\texttt{Jet 1}), where jets are clustered with different jet radii $R=0.3, 0.5, 1.0$.} 
    \label{fig:tQCD_diffr}
\end{figure}

The effects of reclustering the $R = 0.5$ jets with $R = 0.3, 1.0$ are shown in~\cref{fig:tQCD_diffr}. \Cref{fig:tQCD_diffr_a} and~\ref{fig:tQCD_diffr_b} show the number of jets per event when the radius is $0.3$ and $1.0$, respectively, while \cref{fig:tQCD_diffr_c} shows the jet mass distribution of the highest $p_T$ jet for $R = 0.3, 0.5, 1.0$. These distributions are consistent with the fact that intra-jet contributions play a significant role in the classification task for $R = 0.5$ and $R = 1.0,$ and less of a role for $R = 0.3$.

\section{BSM vs. QCD Events} \label{app:details_bsmqcd}

\begin{figure}[h]
    \centering
    \includegraphics[width=0.495\textwidth]{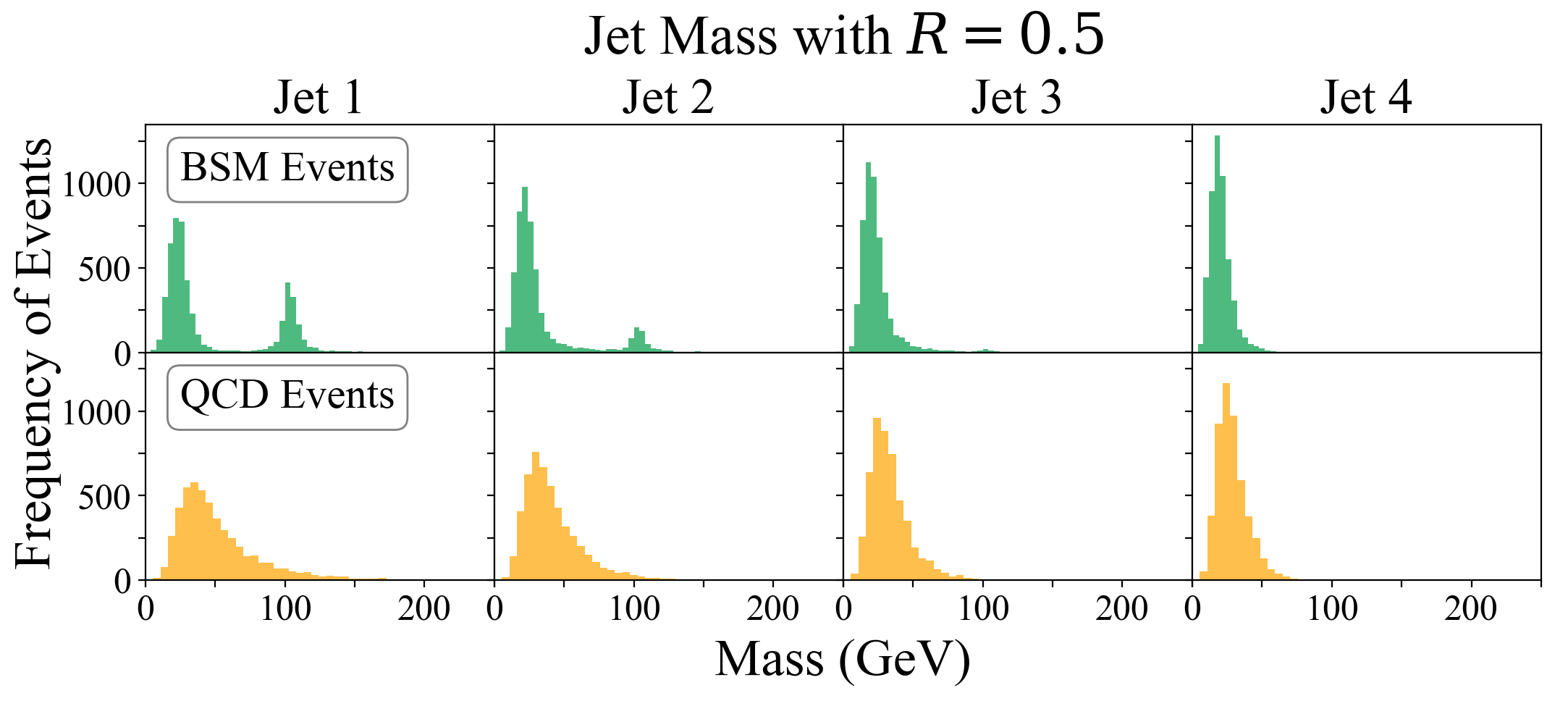}
    \caption{Mass for the first four highest $p_T$ jets (clustered with $R=0.5$) in the signal BSM events (\textit{top}) and the background QCD events (\textit{bottom}).}
    \label{fig:BSMQCD_jetmass_r0.5}
\end{figure}

\Cref{fig:BSMQCD_jetmass_r0.5} again plots the jet mass distribution of the first four highest $p_T$ jets in the BSM signal events and QCD background events. Both \texttt{Jet 1} and \texttt{Jet 2} show a noticeable peak around $m_\phi = 100$ GeV, but are dominated by pure QCD jets. In comparison, the first two jets in the $t \bar{t}$ events have additional information coming from the easily identifiable $W$ boson jets, and include far fewer pure QCD jets. This is consistent with the observation that both intra-jet and inter-jet distributions play a role in the BSM vs.~QCD classification task with $R = 0.5$ jets. 

\begin{figure}[h]
    \begin{subfigure}{0.232\textwidth}
    \includegraphics[width=\textwidth]{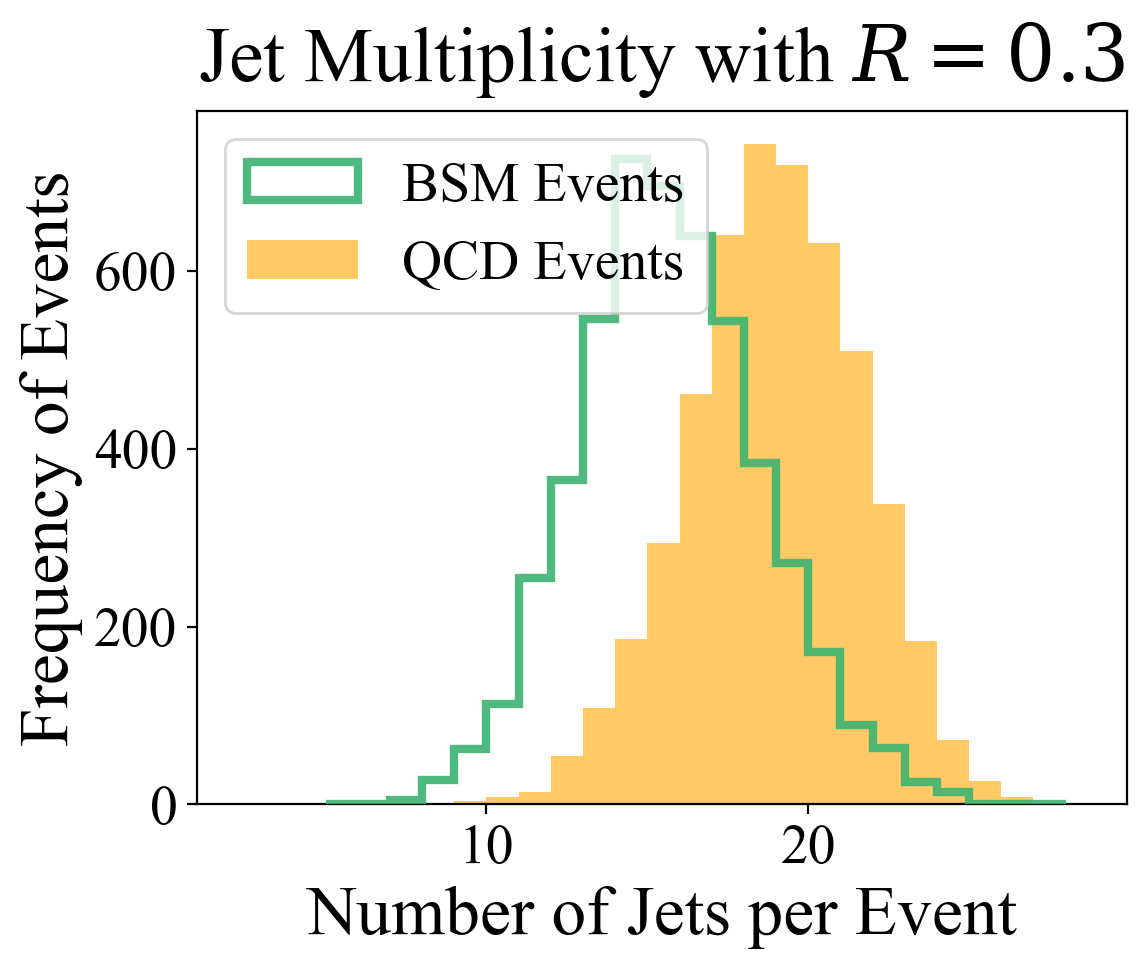}
    \caption{} \label{fig:BSMQCD_diffr_a}
    \end{subfigure}
    \hspace*{\fill} 
    \begin{subfigure}{0.238\textwidth}
    \includegraphics[width=\textwidth]{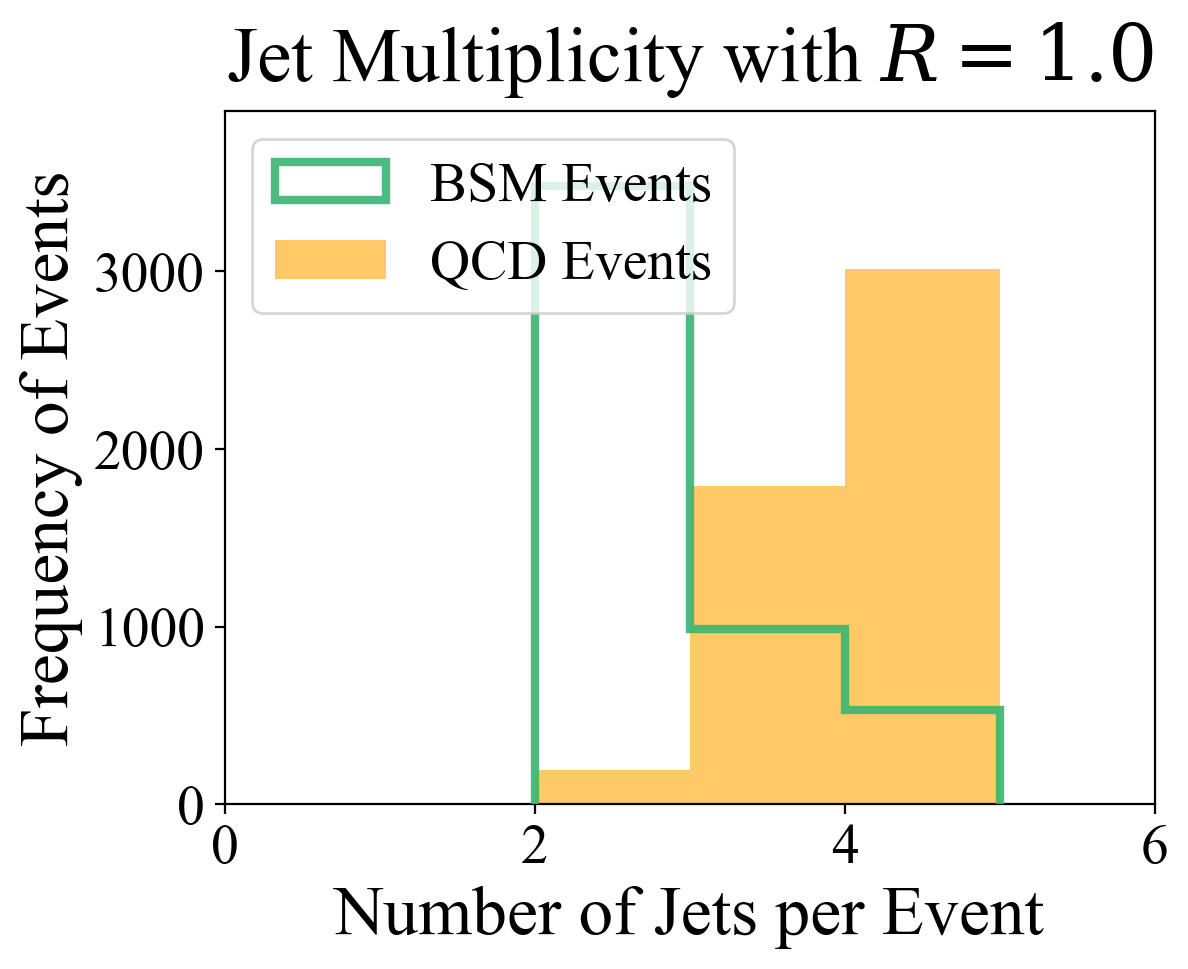}
    \caption{} \label{fig:BSMQCD_diffr_b}
    \end{subfigure}\\
    \begin{subfigure}{0.5\textwidth}
    \includegraphics[width=\textwidth]{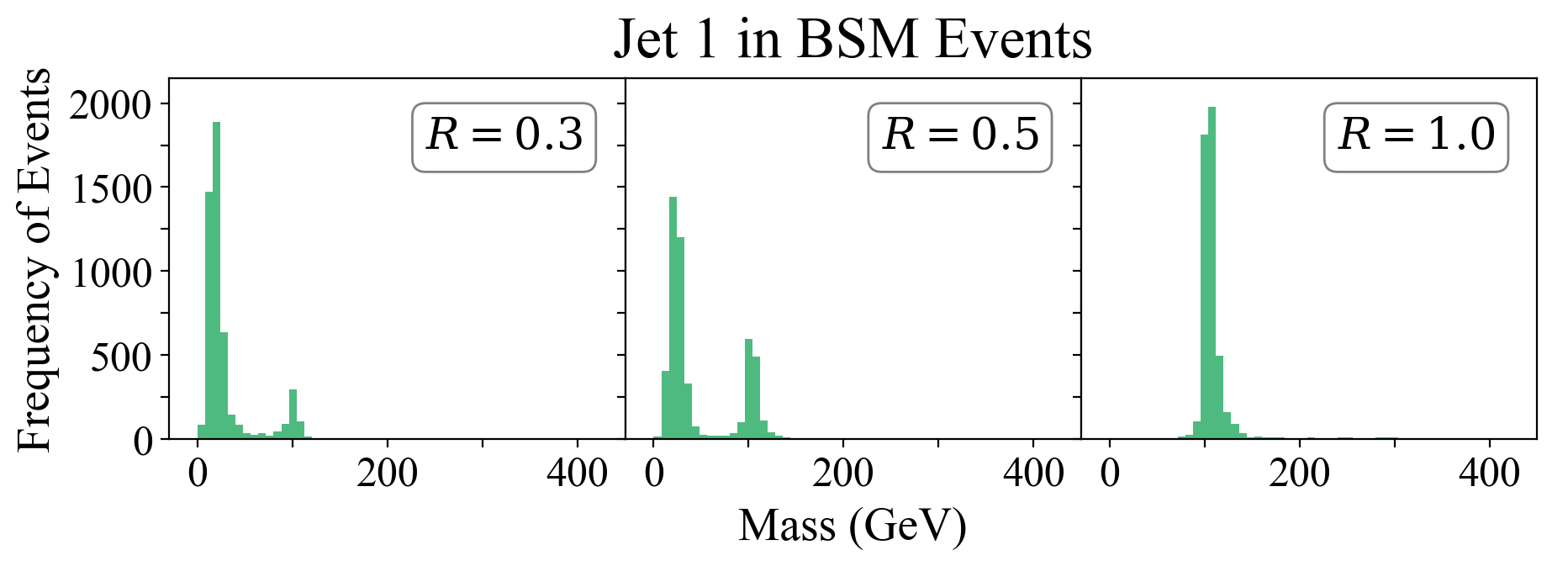}
    \caption{} \label{fig:BSMQCD_diffr_c}
    \end{subfigure}
    \caption{Jet multiplicity in each event, where jets are clustered with a radius of (a) $R=0.3$, or (b) $R=1.0$. (c) Jet mass for the first largest $p_T$ jet in the signal BSM events (\texttt{Jet 1}), where jets are clustered with different jet radii $R=0.3, 0.5, 1.0$.} 
    \label{fig:BSMQCD_diffr}
\end{figure}

The effects of reclustering the $R = 0.5$ jets with $R = 0.3, 1.0$ are shown in~\cref{fig:BSMQCD_diffr}. For $R=1.0$, the decay products of $\phi$ are all clustered together, and the first jet in the signal events is almost always a BSM jet. This is consistent with the improved performance of the ``Jet 1'' multi-scale method for $R = 1.0$ jets. On the other hand, for both $R=0.3$ and $0.5$, the vast majority of \texttt{Jet 1} are QCD jets, consistent with the greater role of inter-jet contributions in the classification task.

\bibliography{draft}

\end{document}